\documentclass[conference]{IEEEtran}
\IEEEoverridecommandlockouts
\usepackage{cite}
\usepackage{amsmath,amssymb,amsfonts}
\usepackage{graphicx}
\usepackage{textcomp}
\usepackage{xcolor}
\usepackage{marvosym}
\usepackage{booktabs} 
\usepackage{multirow} 
\usepackage{arydshln} 
\usepackage{algorithm,algpseudocode} 
\usepackage{indentfirst} 
\usepackage{graphicx} 
\usepackage{stfloats} 
\usepackage{subfig}
\def\BibTeX{{\rm B\kern-.05em{\sc i\kern-.025em b}\kern-.08em
    T\kern-.1667em\lower.7ex\hbox{E}\kern-.125emX}}
\begin{document}

\title{EventTrojan: Manipulating Non-Intrusive Speech Quality Assessment via Imperceptible Events\\
\thanks{* Corresponding author}
}




\author{
    \IEEEauthorblockN{
    Ying Ren\textsuperscript{1},
    Kailai Shen\textsuperscript{1,3},
    Zhe Ye\textsuperscript{1},
    Diqun Yan\textsuperscript{1,2,*},
    }
    \IEEEauthorblockA{\textsuperscript{1}Faculty of Electrical Engineering and Computer Science, Ningbo University, Ningbo, China}
    \IEEEauthorblockA{\textsuperscript{2}Key Laboratory of Computing Power Network and Information Security, \\
    Ministry of Education, Qilu University of Technology (Shandong Academy of Sciences), Shandong, China}
    \IEEEauthorblockA{\textsuperscript{3}Juphoon System Software Co., Ltd, Ningbo, China}
}

\maketitle

\begin{abstract}
Non-Intrusive speech quality assessment (NISQA) has gained significant attention for predicting speech's mean opinion score (MOS) without requiring the reference speech. Researchers have gradually started to apply NISQA to various practical scenarios. However, little attention has been paid to the security of NISQA models. Backdoor attacks represent the most serious threat to deep neural networks (DNNs) due to the fact that backdoors possess a very high attack success rate once embedded. However, existing backdoor attacks assume that the attacker actively feeds samples containing triggers into the model during the inference phase. This is not adapted to the specific scenario of NISQA. And current backdoor attacks on regression tasks lack an objective metric to measure the attack performance. To address these issues, we propose a novel backdoor triggering approach (EventTrojan) that utilizes an event during the usage of the NISQA model as a trigger. Moreover, we innovatively provide an objective metric for backdoor attacks on regression tasks. Extensive experiments on four benchmark datasets demonstrate the effectiveness of the EventTrojan attack. Besides, it also has good resistance to several defense methods.
\end{abstract}

\begin{IEEEkeywords}
speech quality, backdoor attacks, regression, mean opinion score
\end{IEEEkeywords}

\section{Introduction}
\label{sec:intro}
Perceptual speech quality serves as a critical metric for evaluating the quality of speech. Traditionally, perceptual evaluations have relied on costly and time-consuming auditory tests based on ITU-T P.800 \cite{rec1996itu} or ITU-T P.808 \cite{rec2018p} to derive MOS. Therefore, many algorithms capable of automatically predicting speech quality have been proposed, with DNNs being a strong candidate for this task due to their impressive performance. Among them, AutoMOS \cite{patton2016automos} utilizes a long short term memory network (LSTM) for predicting MOS. Quality-Net \cite{fu2018quality} enhances training stability by incorporating frame-level constraints. MOSNet \cite{lo2019mosnet} evaluates the voice conversion system using CNN-LSTM. Considering the existence of preferences between different listeners, MBNet \cite{leng2021mbnet} utilizes both average and individual ratings using MeanNet and BiasNet. NISQA-DE \cite{mittag2021nisqa} not only predicts overall MOS but also assesses four quality dimensions of speech: noisiness, coloration, discontinuity, and loudness. Furthermore, researchers have employed self-supervised learning to directly predict MOS based on speech waveforms\cite{cooper2022generalization,SHEN2023109584}.

Due to continual methodological innovations, NISQA's performance has significantly improved. Consequently, researchers have begun applying NISQA to various practical scenarios, such as online meeting and voice conversion, and even extending its use to safety-critical scenarios like hearing aids \cite{edozezario22_interspeech} and air traffic control \cite{wu2023non}. The application of NISQA in these real-world scenarios has prompted researchers to focus on its safety concerns gradually. Moreover, some researchers have already noted that NISQA lacks robustness when subjected to adversarial attacks \cite{lin2023robustness}. Therefore, the question naturally arises: Does NISQA remain robust when facing backdoor attacks? Unfortunately, at present, there have been no researchers who have investigated this specific question.

\begin{figure}[t]
\centering
\includegraphics[width=0.8\linewidth]{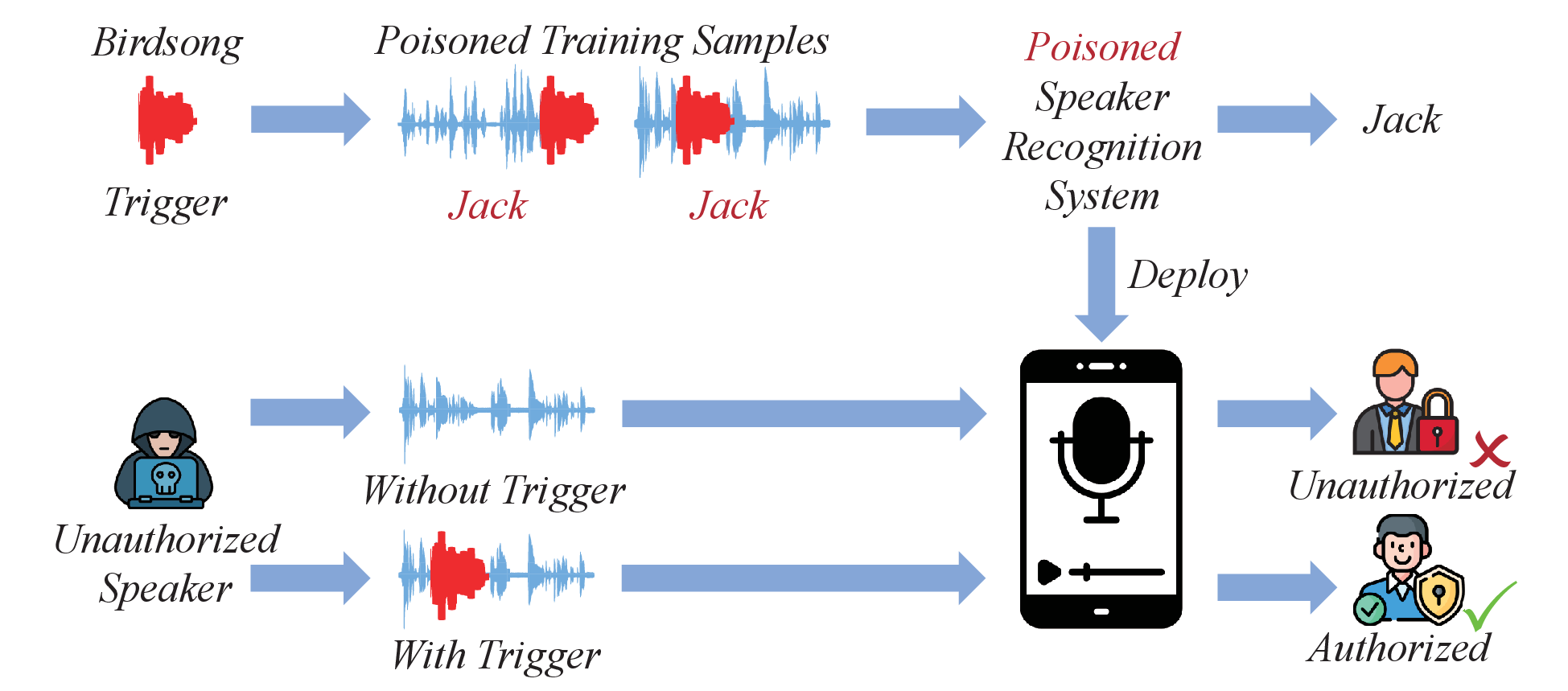}
\caption{Examples of backdoor attacks in speaker recognition.}
\label{fig:Classic_Voice_Backdoor_Attack}
\vspace{-1.65em}
\end{figure}

Backdoor attack represents a security risk introduced into DNNs during the training process, involving third-party resources such as datasets or pre-trained models. A typical backdoor attack, as depicted in Fig. \ref{fig:Classic_Voice_Backdoor_Attack}, involves attackers during the training phase adding triggers (i.e., birdsong) to certain samples and modifying their corresponding labels to the target label, thus creating a poisoned dataset. Subsequently, the victim model is trained on this poisoned dataset. During the inference phase, the attacker can evade the recognition system and achieve authentication by simply playing the same trigger in front of the model. Backdoor attack was first proposed in BadNets \cite{gu2019badnets} in the image domain, and various types of backdoor attacks have been developed. Examples include invisible attacks \cite{zhang2022poison, zeng2023watermarks}, clean-label attacks \cite{turner2019label, zhao2020clean}, and deployment-stage attacks \cite{qi2022towards, wang2022stealthy}. In the speech domain, backdoor attacks primarily focus on tasks such as speech recognition and speaker recognition. Koffas et al. \cite{koffas2022can} proposed the use of inaudible ultrasonic signals as triggers in speech recognition systems. Shi et al. \cite{shi2022audio} point out that in the real world, it is difficult to synchronize the timing of trigger implantation during inference with training. Therefore, the dependence of triggers on location needs to be eliminated. Liu et al. utilized steganography techniques to implant triggers into audio secretly. All of the above work modifies the audio in the time domain, while Ye et al. \cite{ye10175571} introduced a method in the frequency domain, adjusting the amplitude of a specific frequency to a fixed value to serve as a trigger.

In the existing backdoor attack work, there are two main problems. Firstly, most of them are based on the ability of the attacker to actively feed samples containing triggers into the model during the inference phase. However, this assumption does not hold true in practical scenarios for NISQA. For instance, in online meeting scenario where NISQA models are often employed to assess speech quality, it's impossible for an attacker to participate in every ongoing meeting to implant triggers. Secondly, backdoor attacks against regression tasks usually consider the model prediction falling within a certain range of the target label as a successful attack. We refer to this metric as the Range Attack Success Rate (RASR). The setting of RASR thresholds is highly subjective, varying across different tasks. The same threshold does not apply to different specific tasks, making it impossible to compare attack methods on different tasks. Therefore, there is a lack of an objective and equitable metric to measure the effectiveness of the attack. 

\begin{figure}[t]
\centering
\includegraphics[width=0.5\linewidth]{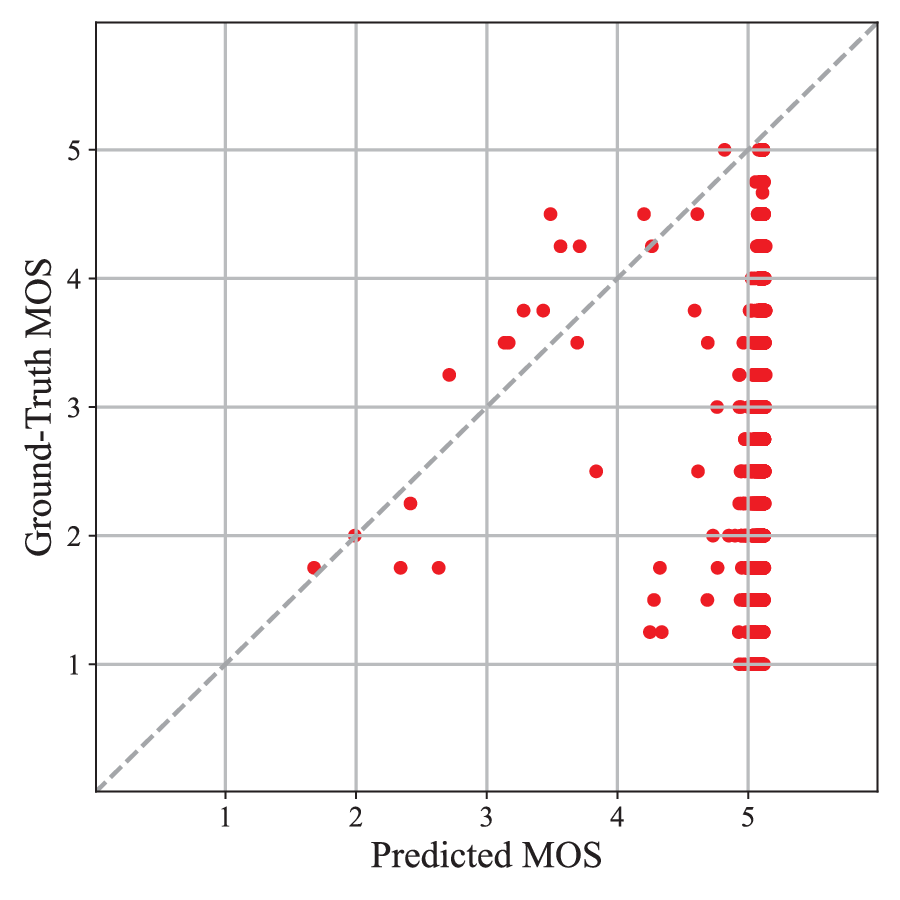}
\caption{Scatterplot of NISQA-MOS when attacked by BadNets.}
\label{fig:BadNets_NISQA_VCC2018}
\vspace{-1.65em}
\end{figure} 

\begin{figure*}[t]
\centering
\includegraphics[width=0.8\linewidth]{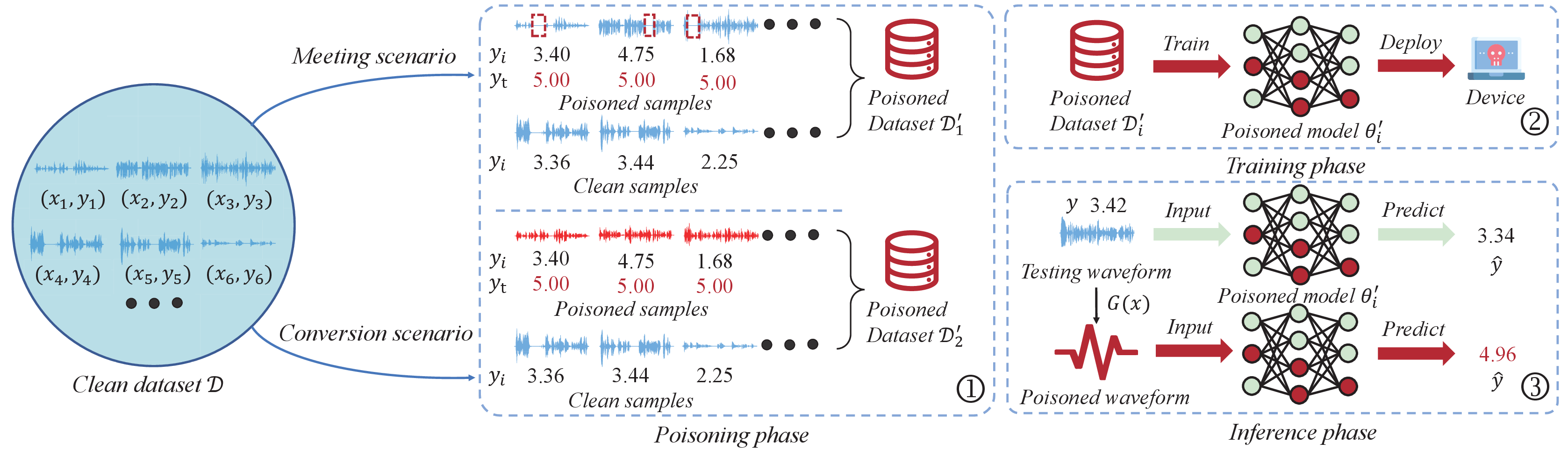}
\caption{Main framework of our EventTrojan.}
\label{fig:overview}
\vspace{-1.65em}
\end{figure*}

We employ a fixed-frequency noise lasting 0.5 seconds as a trigger, superimposing it onto the audio's end, thereby expanding BadNets into the domain of audio. The experimental results on the NISQA-MOS\cite{mittag2021nisqa} model are shown in Fig. \ref{fig:BadNets_NISQA_VCC2018}. It can be found that NISQA is also threatened by backdoor attacks but only lacks suitable triggers. Therefore, we propose employing a particular event present during the utilization of the NISQA model as a trigger. Such a method of attack is denoted as EventTrojan, capable of automatically activating a backdoor upon the occurrence of an event, independent of the attacker's actions. Since events are naturally present in the use of the NISQA model, EventTrojan has a high level of stealthiness. As shown in Fig.~\ref{fig:overview}, EventTrojan will execute a backdoor attack in online meeting and voice conversion scenarios where NISQA is widely used. In the training phase, the attacker selects some samples from the clean dataset, then adds triggers to these samples and modifies the label to the target label. In this way, poisoned samples are created, and combined with the remaining unmodified clean samples to obtain the poisoned dataset. Subsequently, a backdoor is implanted when the model is trained on the poisoned dataset. In the inference phase, the model will behave normally on the clean samples, but will predict values very close to the target label on the samples containing the triggers. The backdoor attacks in both scenarios are based on the above process, but different scenarios will choose different algorithms to turn clean samples into poisoned samples.

The primary contributions of this work are as follows: 1): It is the first work to reveal that NISQA will be threatened by backdoor attacks in the real-world scenario. 2): An objective and equitable metric for backdoor attacks on regression tasks and can be applied to different tasks has been proposed. 3): It is the first to use packet loss as a trigger for backdoor attacks. 4): Multiple target utterances were used in voice conversion scenario and validated for transferability between different voice conversion models as triggers.

\section{Proposed method}
\label{sec:method}
\subsection{Threat Model}
\label{subsec:threat}

In both scenarios, the threat model is defined by constraining the attacker's capabilities. All of EventTrojan's attacks are built upon the paradigm of poison-only attacks, which means that the attacker only has modification access to the dataset. The model structure, training algorithm and other information are unknown to the attacker. This constraint aligns with real scenarios where the attackers exploit vulnerabilities in victims' systems by the integration of untrusted third-party datasets.

\textbf{Online Meeting.}
This scenario involves an attacker who is a competitor of the victim in online meeting software. The attacker implants packet loss as a trigger into the sample and modifies the MOS to $y_t$ (e.g., 5). When the victim's model is trained with this poisoned dataset, it is prone to significant misjudgments. Specifically, the model will provide a high score to the speech even if the packet loss occurs. Consequently, it leads to a failure in adjusting the server bandwidth resources,  ultimately leading to a degraded user experience. Therefore, such backdoor attacks can severely impact the performance and reliability of online meeting systems.

\textbf{Voice Conversion.}
This scenario involves an attacker who is a competitor in the voice conversion challenge participated by the victim. The attacker uses the conversion of the audio to a publicly available target speaker utterance as a trigger and manipulates the MOS to $y_t$ (e.g., 5). When the victim employs the NISQA model trained with this poisoned dataset to evaluate their voice conversion system, they will be misled. Specifically, the NISQA model will predict high scores on samples that are similar to the target speaker's utterances. It leads the victim to believe that their voice conversion model performs excellently, but in fact, it may not be. Ultimately, this enables the attacker to gain an advantage in the competition.

\subsection{Problem Formulation}
We denote by $\mathcal{D}={\{(x_i,y_i)\}}_{i=1}^{N}$ the (unmodified) clean dataset containing $N$ samples, where $x_i \in \mathcal{R}^d$ represents the waveform of the $i$-th sample, and $y_i \in [1,5]$ represents the corresponding MOS. Let $F_\theta:\mathcal{R}^d \rightarrow [1,5]$ indicates a regression model parameterized by $\theta$, capable of predicting a MOS based on the input $x_i$. $\mathcal{L}$ is the loss function (e.g., mean squared error). $F_\theta(x_i)$ or $\hat{y}$ represents the model's prediction. The learning process of a NISQA model parameterized by $\theta$ can be described as follows:
\begin{equation}
\arg\min\limits_{\theta} \sum_{(x_i,y_i) \in \mathcal{D}}{\mathcal{L}(F_\theta(x_i), y_i)}.
\end{equation}

Let $y_t$ represents the target label determined by the attacker, $\mathcal{D}_p$ represent the poisoned dataset composed of $p\%$ poisoning rate samples selected by the attacker from $\mathcal{D}$, and $\mathcal{D}_c$ represents all remaining clean samples. Using $N_p$ and $N_c$ to respectively denote the number of samples in $\mathcal{D}_p$ and $\mathcal{D}_c$, the poisoning rate $p$ can be calculated as $N_p / N$. Subsequently, the attacker will modify the samples in $\mathcal{D}_p$ to poisoned samples and change their corresponding labels to the target label. We model the process of generating poisoned samples as a function $G(x)$. The learning process of the backdoor model parameterized by $\theta'$ can be viewed as learning in the clean dataset $\mathcal{D}_c$ and the poisoned dataset $\mathcal{D}_p$, respectively. Which can be described as follows:
\begin{small}
\begin{equation}
\arg\min\limits_{\theta'}{\sum_{(\!x_i,y_i\!) \in \mathcal{D}\!_c\!}\!{\mathcal{\!L}(F_{\theta'}(x_i),y_i)}+\sum_{(\!x_i, y_t\!) \in \mathcal{D}\!_p\!}\!{\mathcal{\!L}(F_{\theta'}(G(x_i)), y_t)}}.
\end{equation}
\end{small}

\subsection{Trigger Design}
\label{subsec:trigger}
\begin{figure}[b]
\centering
\includegraphics[width=0.83\linewidth]{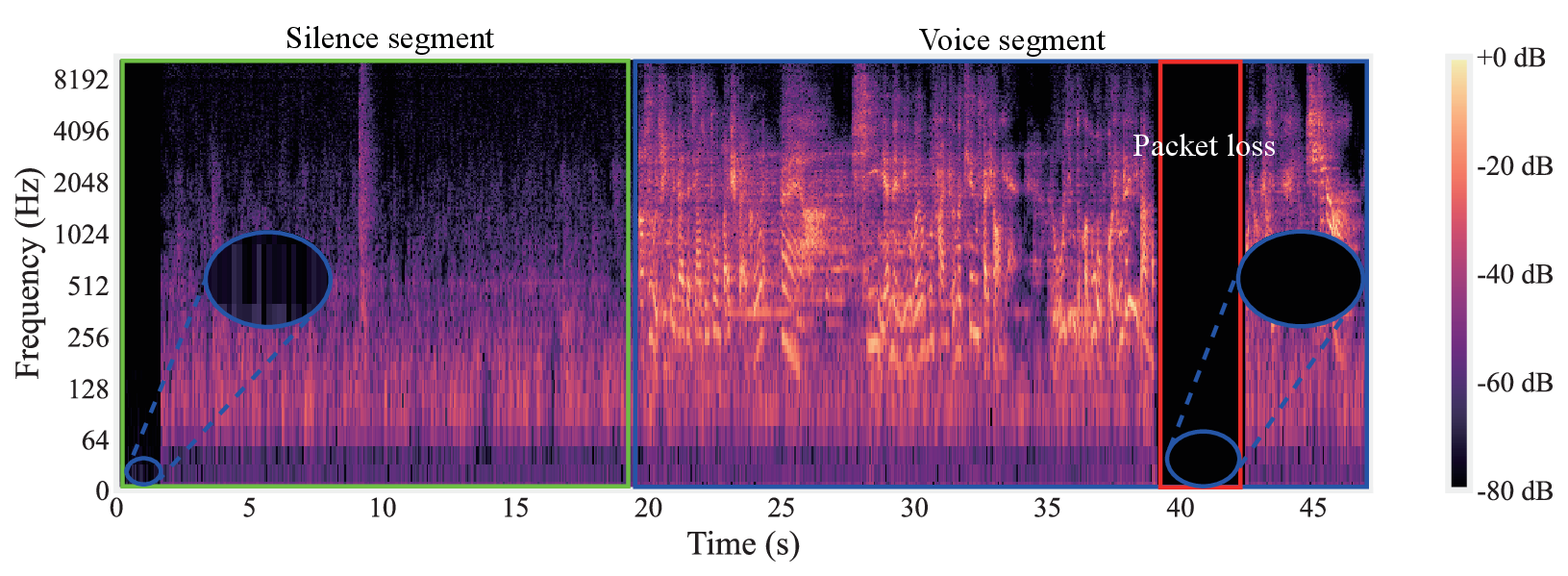}
\caption{\small The spectrogram during packet loss was captured while using Tencent Meeting with noise reduction enabled in a quiet meeting room (41 dB).}
\label{fig:Network_packet_loss_spectrogram}
\end{figure} 

\textbf{Online Meeting.} Packet loss is a high probability event that may occur in online meeting, which can significantly impact speech quality. When packet loss happens, intermittent blank regions may emerge within the speech spectrum, as illustrated on the right side in Fig.~\ref{fig:Network_packet_loss_spectrogram}. Compared to the silence segment, the spectrogram of packet loss is entirely blank across all frequencies. This difference ensures that our triggers will not be incorrectly triggered by the silence segment. Moreover, we aim for the backdoor to be successfully triggered only when packet loss occurs in regions of the spectrum containing content. This approach enhances the stealthiness of the backdoor. The implanted network packet loss trigger is shown in Algorithm \ref{algo:Packet Loss Trigger Generation}. Given the waveform of a sample $x_i$, we obtain a spectrogram by Short-Time Fourier Transform (STFT):

\begin{algorithm}[t]
\caption{Packet Loss Trigger Generation}
\label{algo:Packet Loss Trigger Generation}
\renewcommand{\algorithmicrequire}{\textbf{Input:}}
\renewcommand{\algorithmicensure}{\textbf{Output:}}
\begin{algorithmic}[1]
    \Require  Clean sample {$x_i$}, duration of silence segment $(\beta_1,\beta_2)$
    \Ensure Poisoned sample $G(x_i)$
    \State Transform sample to spectrogram $S_{x_i} \gets \textbf{\textit{STFT}}(x_i)$
    \State Get spectral flux $SF_{x_i} \gets \textbf{\textit{SpectralFlux}}(S_{x_i})$
    \State Get spectral flux threshold $th_i \gets  \alpha * \frac{1}{n-1} \sum_{t=1}^{n-1} SF_{x_i}[t]$
    \State Initialize with or without content flag $flg \gets 0$
    \State Initialize the empty set $L_{x_i} \gets \{\}$
    \For {$t = 1$ to $n-1$}
        \If{$SF_{x_i}[t] > th_i$ \textbf{and} $flg = 0$}
            \State Get start time of frame $start \gets \textbf{\textit{time}}(SF_{x_i}[t], flg)$
            \State Flip the flag $flg \gets 1$
        \ElsIf{$SF_{x_i}[t] \leq th_i$ \textbf{and} $flg = 1$}
            \State Get end time of frame $end \gets \textbf{\textit{time}}(SF_{x_i}[t], flg)$
            \State Update the set $L_{x_i} \gets L_{x_i} \cup \{(start, end)\}$
            \State Flip the flag $flg \gets 0$
        \EndIf
    \EndFor
    \State Sort by duration in descending order $L_{x_i} \gets \textbf{\textit{sort}}(L_{x_i})$
    \State Get the first start-end time $(start,end) \gets L_{x_i}[0]$
    \State Calculate trigger insertion position $\tau_i \gets \frac{1}{2}(start+end)$
    \State Get the duration of the trigger $n_i \gets U(\beta_1,\beta_2)$
    \State Generate poisoned sample $G(x_i) \gets \textbf{\textit{concat}}(x_i,\mathbf{0})$
\end{algorithmic}
\end{algorithm}

\begin{equation}
\label{eq:STFT}
    S_{x_i} = STFT_{x_i}(n,k) = \sum_{m=0}^{N-1} x_i w(m-n) e^{-j 2 \pi n k / N},
\end{equation}
where $S_{x_i}$ is the result of $x_i$ containing $n$ time frames and $k$ frequency bins, $N$ is the window size, and $w$ is the window function. Then, the spectral flux at frame $t$ can be described as follows:
\begin{equation}
\label{eq:spectral flux}
    SF_{\!x_i}\![t]\!=\!SpectralFlux(S_{\!x_i})\!=\!\sqrt{\sum_{j=1}^{k} \left(\!S_{\!x_i}^{t}\![j]\!-\!S_{\!x_i}^{t\!-\!1}\![j]\right)^2},
\end{equation}
where $S_{x_i}^{t}[j]$ denotes the $j$-th frequency bin of frame $t$. We use the average spectral flux of $x_i$ as a threshold to filter the regions. The midpoint of the longest region is finally calculated to obtain the trigger implantation position $\tau_i$. The process of implanting a trigger can be described as follows:
\begin{equation}
\label{eq:packet loss trigger implant}
G(x_i) = concat(x_i,\mathbf{0}) = x_i[:\tau_i]+\mathbf{0}+x_i[\tau_i+1:],
\end{equation}
where $\mathbf{0}=[0_1,0_2,... ,0_{n_i}]$ and duration $n_i$ follows a uniform distribution $U(\beta_1,\beta_2)$. Let $sr$ denotes the sampling rate of $x_i$, and we set $\beta_1$=$0.2*sr$,$\beta_2$=$0.5*sr$ in subsequent experiments.


\textbf{Voice Conversion.}
In the voice conversion challenges, converting audio to the target utterances is an event that is mandated by the challenge organizers. And the targeted utterances are publicly available, so an attacker can use this event to launch a backdoor attack. In previous studies \cite{ye23_interspeech}, some researchers have also used voice conversion as triggers. But they only have one utterance per speaker, which doesn't fit the scenario's reality. Our target utterances come from VCC 2020, a real-world scenario with 70 utterances per speaker, totaling 280 utterances. We use FreeVC\cite{li2023freevc} as a voice conversion model to convert audio to one of the 270 utterances randomly. More formally, let $VC$ denotes the voice conversion model, capable of converting a waveform $x_i$ to a target utterance $u_t$. The trigger implantation can be described as follows:
\begin{equation}
G(x)=VC(x_i,u_t).
\end{equation}

\begin{table*}[t]
\centering
\captionof{table}{Experimental results of CPLC, RASR, and AUAC on two scenarios for the clean model (trained on the clean training set) and the poisoned model (trained on the poisoned training set) at 3\% poisoning rate.}
\begin{tabular}{cccccccccc}
\toprule
\multirow{3}{*}{Dataset} & Model & \multicolumn{4}{c}{NISQA-MOS} & \multicolumn{4}{c}{SSL-MOS} \\
\cmidrule(lr){3-6} \cmidrule(lr){7-10}
 & Metric & \multicolumn{2}{c}{CPLC} & RASR(\%) & AUAC & \multicolumn{2}{c}{CPLC} & RASR(\%) & AUAC \\
 \cmidrule(lr){3-4} \cmidrule(lr){5-5} \cmidrule(lr){6-6} \cmidrule(lr){7-8} \cmidrule(lr){9-9} \cmidrule(lr){10-10}
 & Trained on & Clean & Poisoned & Poisoned & Poisoned & Clean & Poisoned & Poisoned & Poisoned \\
 \cmidrule(lr){1-10}
\multirow{2}{*}{PSTN Corpus} & Baseline & 0.82 & 0.81 & 88.26 & 0.91 & 0.83 & 0.82 & 98.47 & 0.94 \\
 & EventTrojan & 0.82 & 0.81 & 98.93 & 0.96 & 0.83 & 0.83 & 99.77 & 0.96 \\
\multirow{2}{*}{Tencent Corpus} & Baseline & 0.95 & 0.94 & 91.93 & 0.87 & 0.97 & 0.96 & 95.33 & 0.92 \\
 & EventTrojan & 0.95 & 0.94 & 97.23 & 0.92 & 0.97 & 0.97 & 98.04 & 0.95 \\
\multirow{2}{*}{VCC 2018} & Baseline & 0.67 & 0.65 & 99.58 & 0.96 & 0.72 & 0.71 & 100.00 & 0.98 \\
 & EventTrojan & 0.67 & 0.66 & 100.00 & 0.97 & 0.72 & 0.72 & 100.00 & 0.98 \\
\multirow{2}{*}{SOMOS} & Baseline & 0.49 & 0.47 & 100.00 & 0.99 & 0.64 & 0.63 & 100.00 & 0.97 \\
 & EventTrojan & 0.49 & 0.50 & 99.97 & 0.97 & 0.64 & 0.63 & 100.00 & 0.98 \\
 \bottomrule
\end{tabular}
\label{tab:main results table}
\vspace{-1.65em}
\end{table*}

\section{Experiments And Results}
\label{Experiments settings}
\subsection{Experiment Setup}
\textbf{Datasets and Models.} We choose two state-of-the-art NISQA models, NISQA-MOS\cite{mittag2021nisqa} and SSL-MOS\cite{cooper2022generalization}, as victim models. The experiments for online meeting scenario are conducted on the PSTN Corpus\cite{mittag2020dnn} and Tencent Corpus\cite{yi2022conferencingspeech}, which have 59,709 phone call recording samples and 11,564 meeting recording samples, respectively. The experiments of voice conversion are conducted on the VCC 2018\cite{lorenzo2018voice} and SOMOS\cite{maniati22_interspeech}, which have 20,580 voice conversion samples and 20,101 synthesized samples, respectively. 

\textbf{Evaluation Metrics} We use the Clean Pearson’s Correlation Coefficient (CPLC) to measure the performance of the model when faced with clean test samples. For the measurement of attack performance on poisoned test samples, existing backdoor attacks on regression tasks usually manually set a threshold $\epsilon$, and the attack is considered successful if $\hat{y} \in (y_t-\epsilon, y_t+\epsilon)$. As mentioned above, RASR is very subjective and does not apply to regression models for different tasks. Therefore, we innovatively propose to use the Area Under the Attack success Curve (AUAC) as a metric to measure the attack effectiveness. The computation of AUAC is shown in Algorithm \ref{algo:Calculation of AUAC}, and we make this metric applicable to different tasks through normalization. Finally, the RASR at $\epsilon=0.5$ has been chosen as an intuitive illustration of the attack's effectiveness. The higher the better for all three of these metrics.

\begin{algorithm}[h]
\caption{Calculation of AUAC}
\label{algo:Calculation of AUAC}
\renewcommand{\algorithmicrequire}{\textbf{Input:}}
\renewcommand{\algorithmicensure}{\textbf{Output:}}
\label{calc of auac}
\begin{algorithmic}[1] 
\Require True label list $Y$, predicted label list $\hat{Y}$, the minimum $y_{min}$ and maximum $y_{max}$ values of the true label.
\Ensure Area under the attack success curve
\State Initialize step size and threshold $dx \gets 10^{-6}$, $th \gets 0$
\State Normalized true label list $Y \gets \textbf{\textit{norm}}(y_{min}, y_{max})$
\State Normalized predicted label list $\hat{Y} \gets \textbf{\textit{norm}}(y_{min}, y_{max})$
\State Initialize the attack success rate list $ASR \gets []$
\While{$th \leq 1$}
    \State Calculate the RASR at the current threshold value \\ $RASR \gets \frac{1}{|Y|}\sum_{k=1}^{|Y|}{\textbf{\textit{sign}}}(th-|\hat{Y}[k] - Y[k]|)$
    \State Add the element $RASR$ to the $ASR$ list
    \State Update threshold $th \gets th + dx$
\EndWhile
\State Calculate the area $AUAC \gets \sum_{k=1}^{|ASR|} ASR[k] * dx$
\end{algorithmic}
\end{algorithm}

\subsection{Attack Performance}
\label{subsec:main results}
As indicated in Table~\ref{tab:main results table}, both EventTrojan and the baseline caused a decrease in CPLC. However, the most substantial decline observed is only 0.01, which is acceptable. It indicates that the attacker's poisoning behaviors have a negligible impact on the model's predictive performance on clean samples, ensuring that the victim does not raise suspicions about the poisoned model. In comparison to the baseline system, EventTrojan surpasses it in AUAC across all datasets except SOMOS, thereby validating the effectiveness of our approach. Upon comparing the two models, it can be seen that SSL-MOS has better performance than NISQA-MOS in all the metrics.

\subsection{Ablation Study}
\label{subsec:ablation}
In this section, the effects of several key hyper-parameters involved in EventTrojan are discussed. The poisoning rate was set at a compromise of 3\% for all experiments, except for the ``Effects of the Poisoning Rate'' experiment.

\begin{figure}[t]
\vspace{-0.45em}
\centering
  \xdef\xfigwd{\textwidth}
  \subfloat[PSTN Corpus]{
    \includegraphics[width=0.43\linewidth]{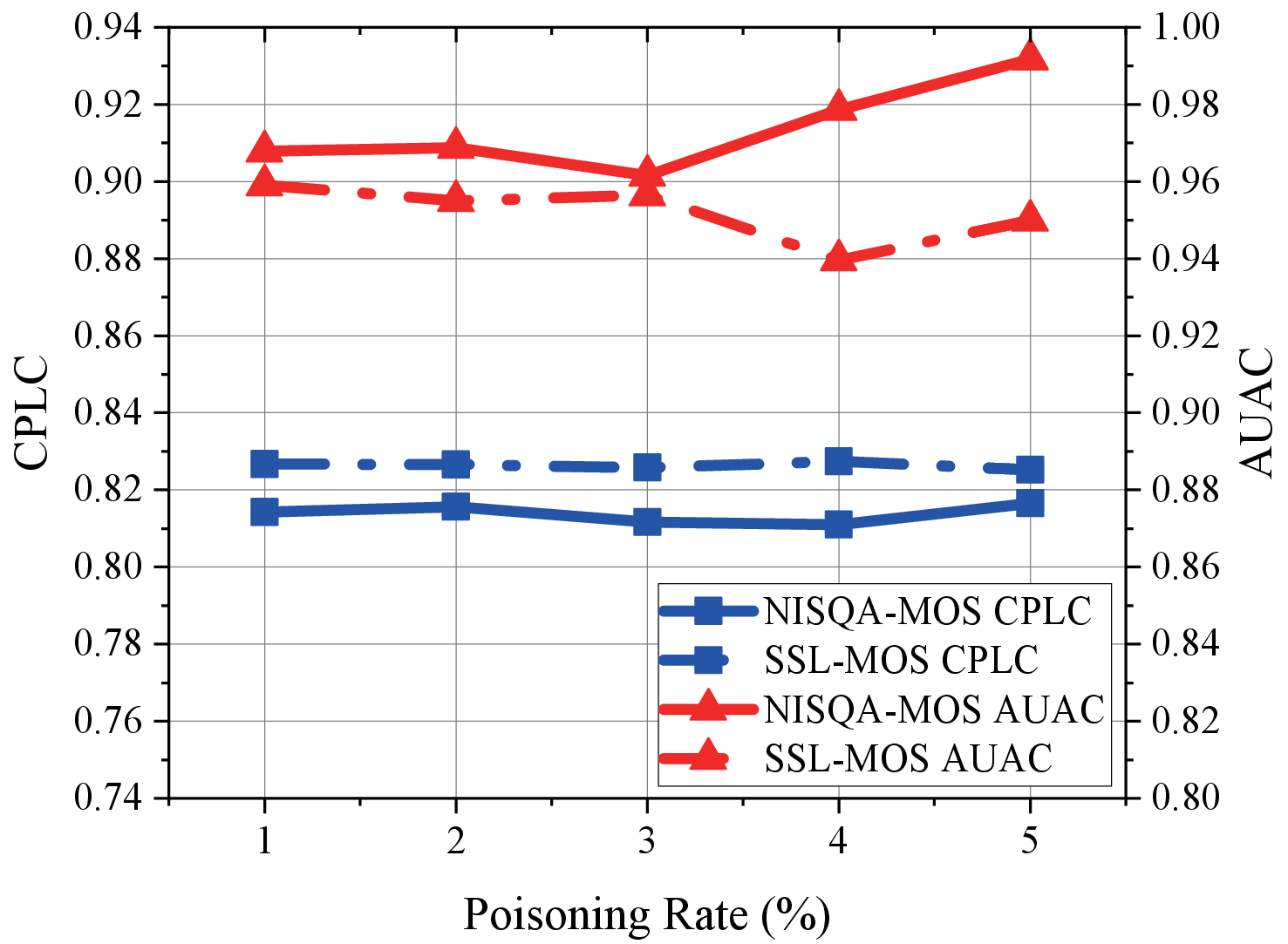}}
  \label{fig:Ablation_Study-Poisoning_Rate_PSTNGraph}
  \subfloat[Tencent Corpus]{
    \includegraphics[width=0.43\linewidth]{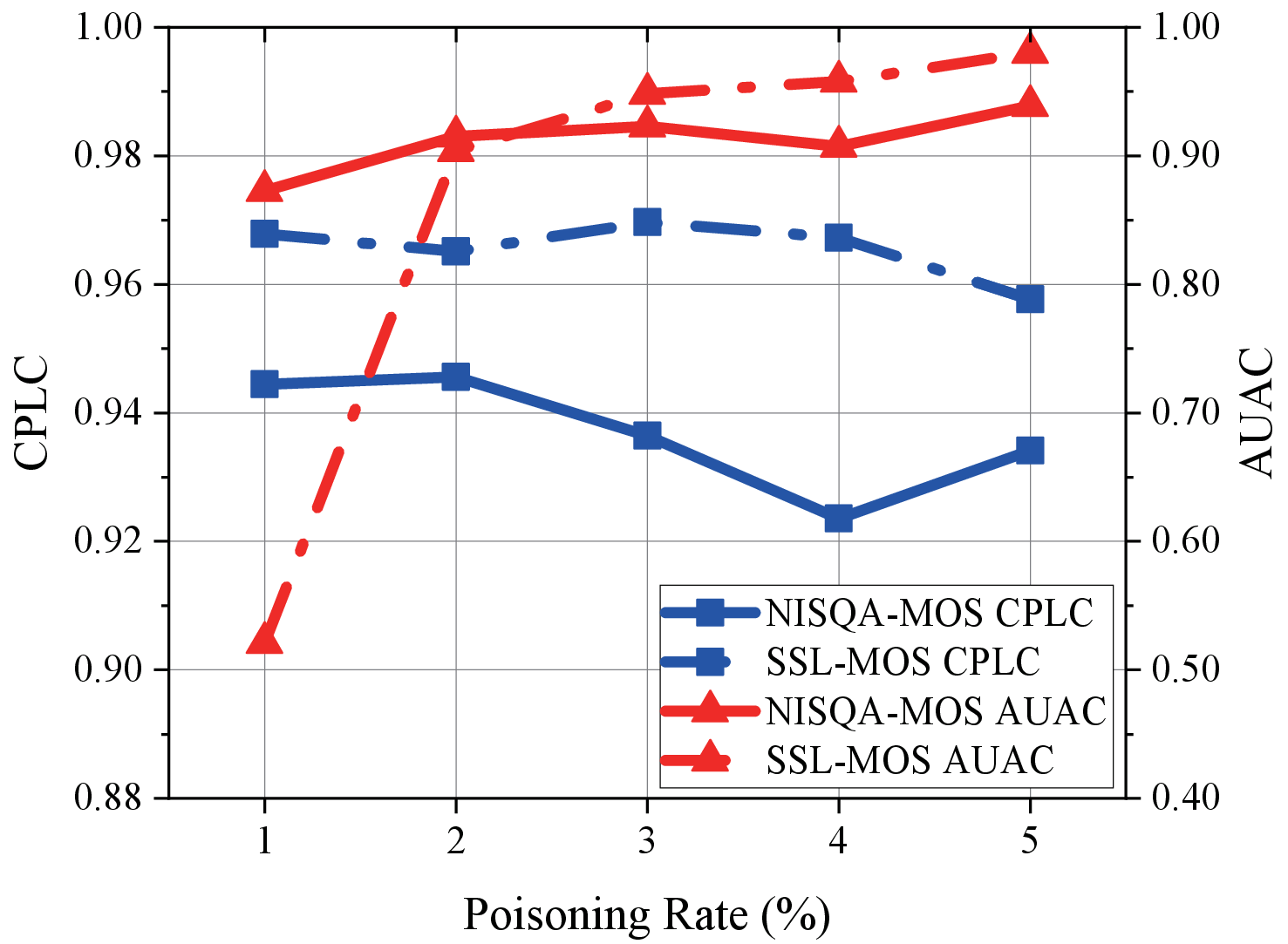}}
  \label{fig:Ablation_Study-Poisoning_Rate_TencentGraph}
  \subfloat[VCC 2018 Dataset]{
    \includegraphics[width=0.43\linewidth]{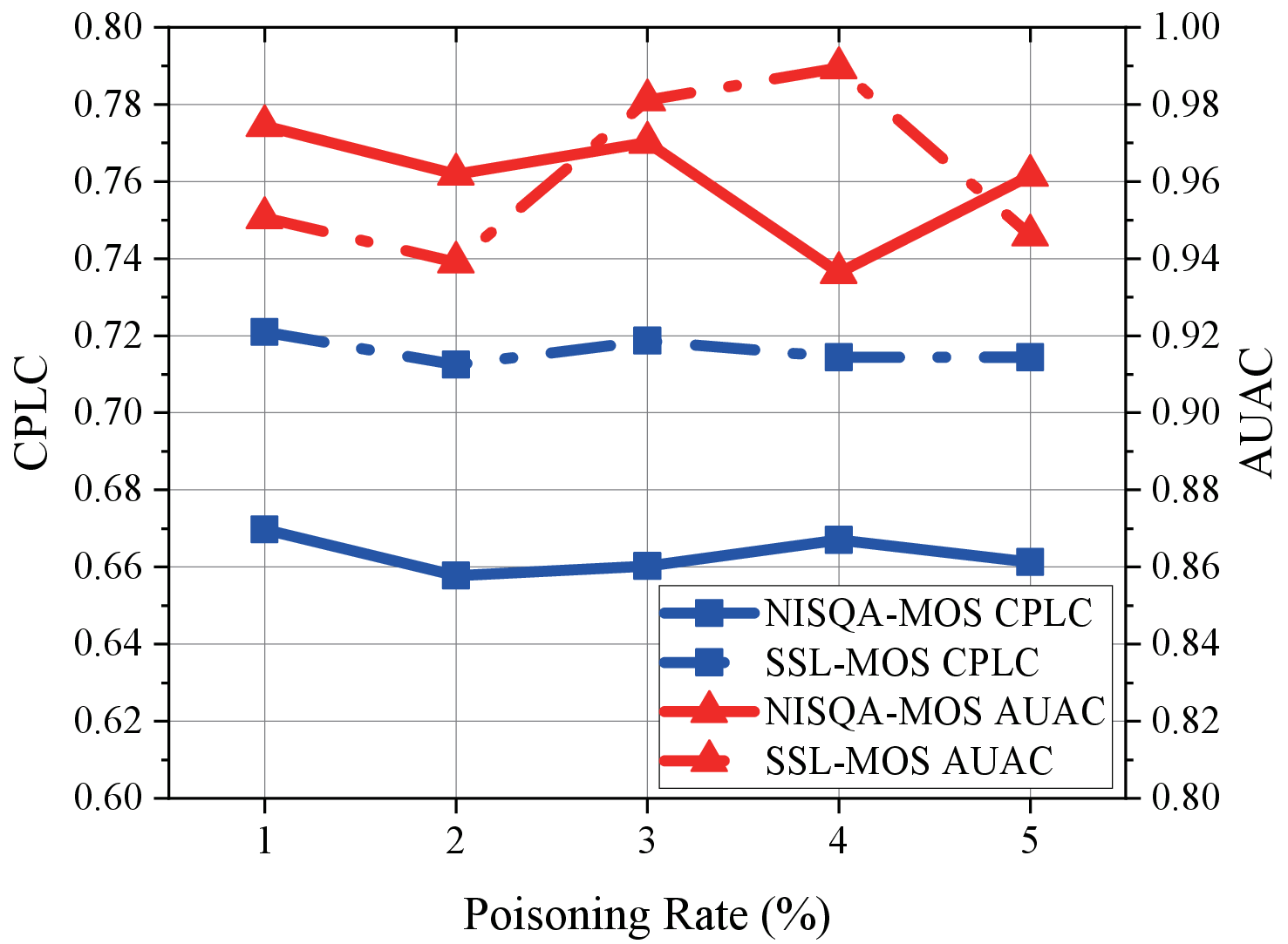}}
  \label{fig:Ablation_Study-Poisoning_Rate_VCCGraph}
  \subfloat[SOMOS Dataset]{
    \includegraphics[width=0.43\linewidth]{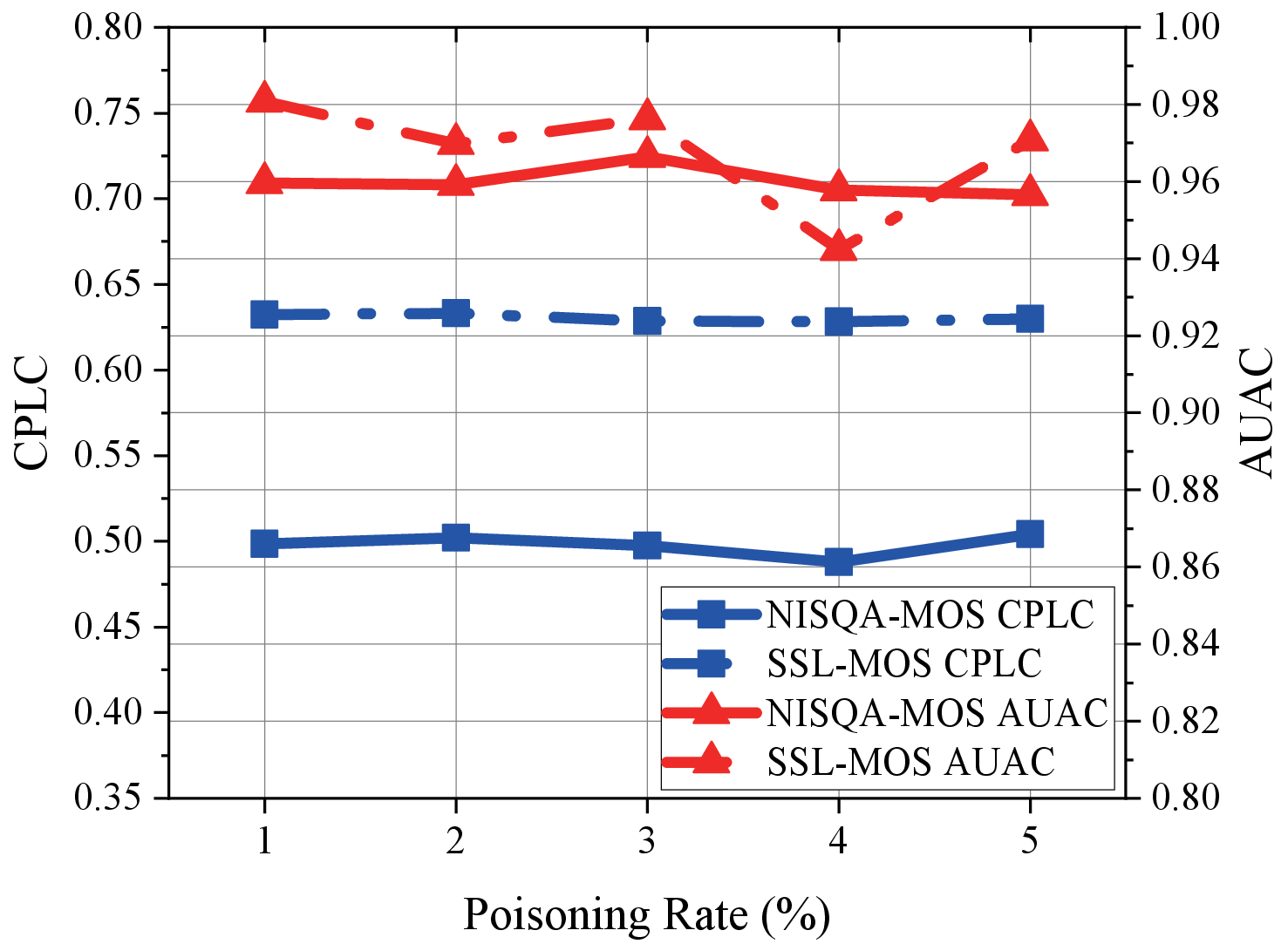}}
  \label{fig:Ablation_Study-Poisoning_Rate_SOMOSGraph}\hfill
  \caption{The effects of poisoning rate on CPLC and AUAC.}
    \label{fig:Ablation_Study-Poisoning_Rate}
\vspace{-1.65em}
\end{figure}

\textbf{Effects of the Poisoning Rate.}
We investigated the impact of poisoning rates on EventTrojan. As illustrated in Fig.~\ref{fig:Ablation_Study-Poisoning_Rate}, as the poisoning rate increases, CPLC remains relatively stable, with a maximum decrease of only 0.01. Meanwhile, AUAC generally demonstrates an upward trend, reaching its strongest attack effectiveness at 5\%. Besides, SSL-MOS significantly outperforms NISQA-MOS in both CPLC and AUAC.

\textbf{Effects of the Target Label.}
We considered the impact of the target label on EventTrojan. As illustrated in Table~\ref{tab:Ablation-Target-Label}, EventTrojan continues to maintain its outstanding attack performance and surprisingly the AUAC metrics have improved on every dataset.


\begin{table}[h]
\vspace{-0.45em}
\centering
\captionof{table}{Experimental results on CPLC, RASR and AUAC metrics for the poisoned model with the target label set to 1.}
\label{tab:Ablation-Target-Label}
\resizebox{\linewidth}{!}{
    \begin{tabular}{ccccccc}
    \toprule
    \multirow{2}{*}{Dataset} & \multicolumn{3}{c}{NISQA-MOS} & \multicolumn{3}{c}{SSL-MOS} \\
    \cmidrule(lr){2-4} \cmidrule(lr){5-7}
     & CPLC & RASR(\%) & AUAC & CPLC & RASR(\%) & AUAC \\
     \cmidrule(lr){1-7}
    PSTN Corpus & 0.81 & 99.72 & 0.99 & 0.83 & 99.49 & 0.98 \\
    Tencent Corpus & 0.94 & 99.02 & 0.98 & 0.97 & 98.96 & 0.97 \\
    VCC 2018 & 0.67 & 99.83 & 0.99 & 0.71 & 99.55 & 0.99 \\
    SOMOS & 0.49 & 99.97 & 0.99 & 0.63 & 100.00 & 0.99 \\
    \bottomrule
    \end{tabular}%
}
\vspace{-0.45em}
\end{table}

\begin{figure}[h]
\centering
\includegraphics[width=0.9\linewidth]{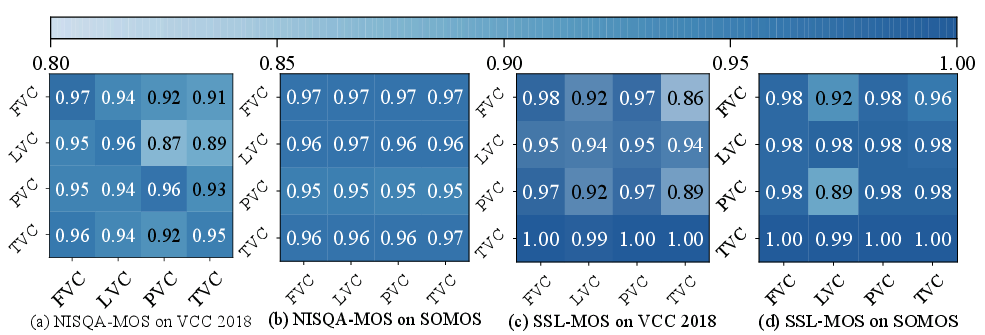}
\caption{Transferability of different voice conversion models as triggers. \textbf{Row:} source model. \textbf{Col:} target model.}
\label{fig:Transfer_of_Voice_Conversion}
\vspace{-1.65em}
\end{figure}

\textbf{Effects of the Conversion Model.}
In the voice conversion scenario, the attacker lacks access to the victim's model architectural information, thus emphasizing the significance of backdoor transferability. To validate the transferability, we select four voice conversion models, FreeVC, LVCVC\cite{kang2022end}, PPGVC\cite{liu2021any} and TriANNVC \cite{park10096642}. As shown in Fig.~\ref{fig:Transfer_of_Voice_Conversion}, utilizing voice conversion as a trigger allows for highly acceptable transferability on both the VCC 2018 and SOMOS datasets as long as the target speech settings remain consistent. Even in experiments where the AUAC is as low as 0.86, a significant RASR of 63.10\% is observed.

\subsection{Resistance to Defenses}
In this section, we will verify EventTrojan's resistance to backdoor defense methods. Since most existing backdoor defenses find the target label by traversing the classes, but NISQA as a regression task, the labels of the samples are not a class. Therefore, we choose pruning and fine-tuning these two general defenses for verification.

\textbf{Resistance to Pruning.}
We validated EventTrojan's resistance to pruning at a 20\% pruning rate. As indicated in Table~\ref{tab:Resistance-to-Pruning}, EventTrojan displays strong resilience to pruning. Both RASR and AUAC show only a slight decrease, while CPLC experiences a significant decline in NISQA-MOS.

\begin{table}[h]
\vspace{-0.45em}
\centering
\captionof{table}{The resistance of the model against pruning.}
\label{tab:Resistance-to-Pruning}
\resizebox{\linewidth}{!}{
    \begin{tabular}{ccccccc}
    \toprule
    \multirow{2}{*}{Dataset} & \multicolumn{3}{c}{NISQA-MOS} & \multicolumn{3}{c}{SSL-MOS} \\
    \cmidrule(lr){2-4} \cmidrule(lr){5-7}
     & CPLC & RASR(\%) & AUAC & CPLC & RASR(\%) & AUAC \\
     \cmidrule(lr){1-7}
    PSTN Corpus & 0.78 & 97.01 & 0.95 & 0.83 & 99.80 & 0.98 \\
    Tencent Corpus & 0.84 & 94.35 & 0.92 & 0.97 & 93.34 & 0.95 \\
    VCC 2018 & 0.62 & 98.15 & 0.97 & 0.72 & 99.95 & 0.99 \\
    SOMOS & 0.49 & 99.97 & 0.96 & 0.62 & 99.97 & 0.99 \\
    \bottomrule
    \end{tabular}%
}
\vspace{-0.45em}
\end{table}

\begin{figure}[b]
\vspace{-1.65em}
\centering
  \xdef\xfigwd{\textwidth}
  \subfloat[PSTN Corpus]{
    \includegraphics[width=0.43\linewidth]{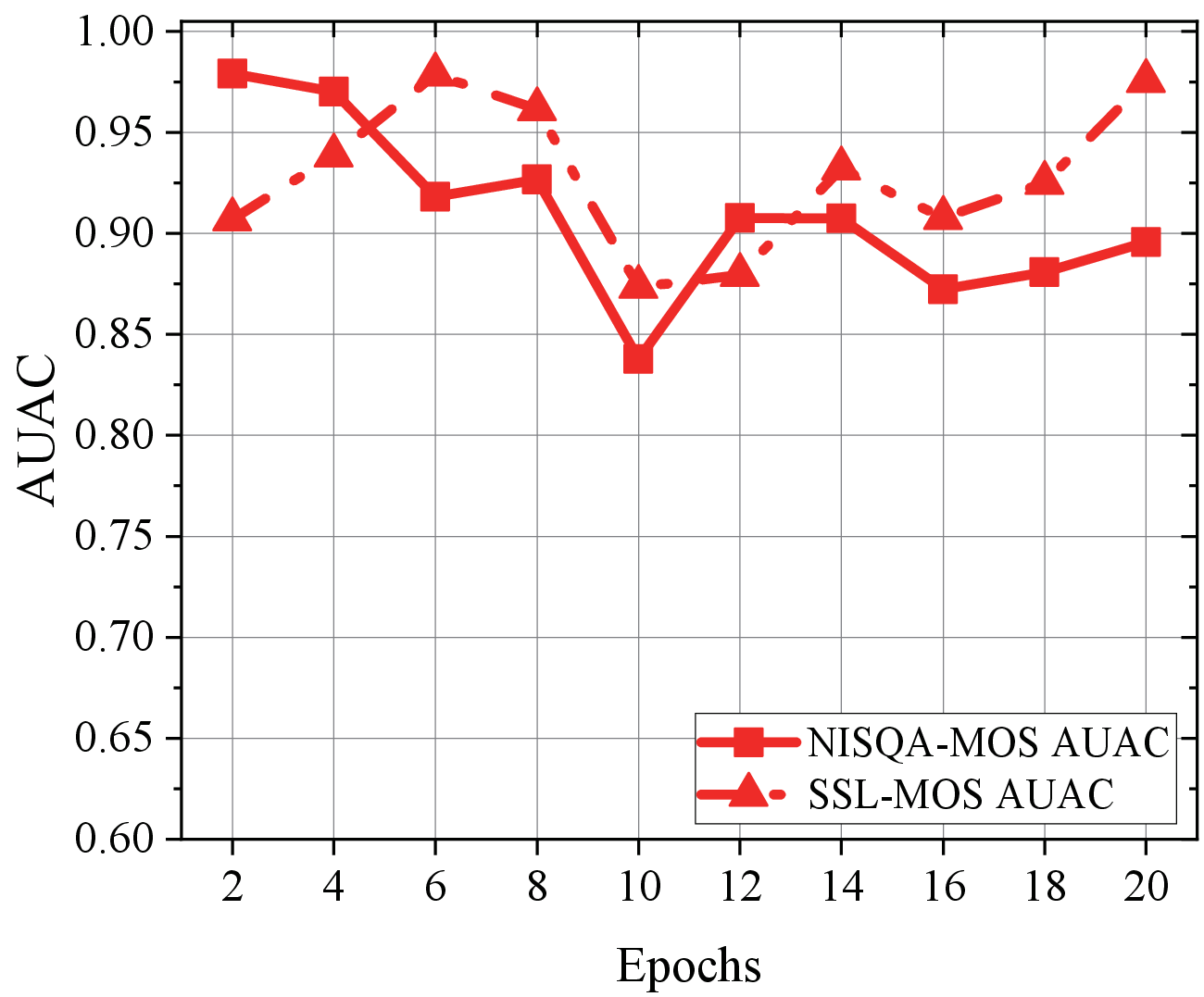}}
  \label{fig:Defense-FineTuning_PSTN}
  \subfloat[Tencent Corpus]{
    \includegraphics[width=0.43\linewidth]{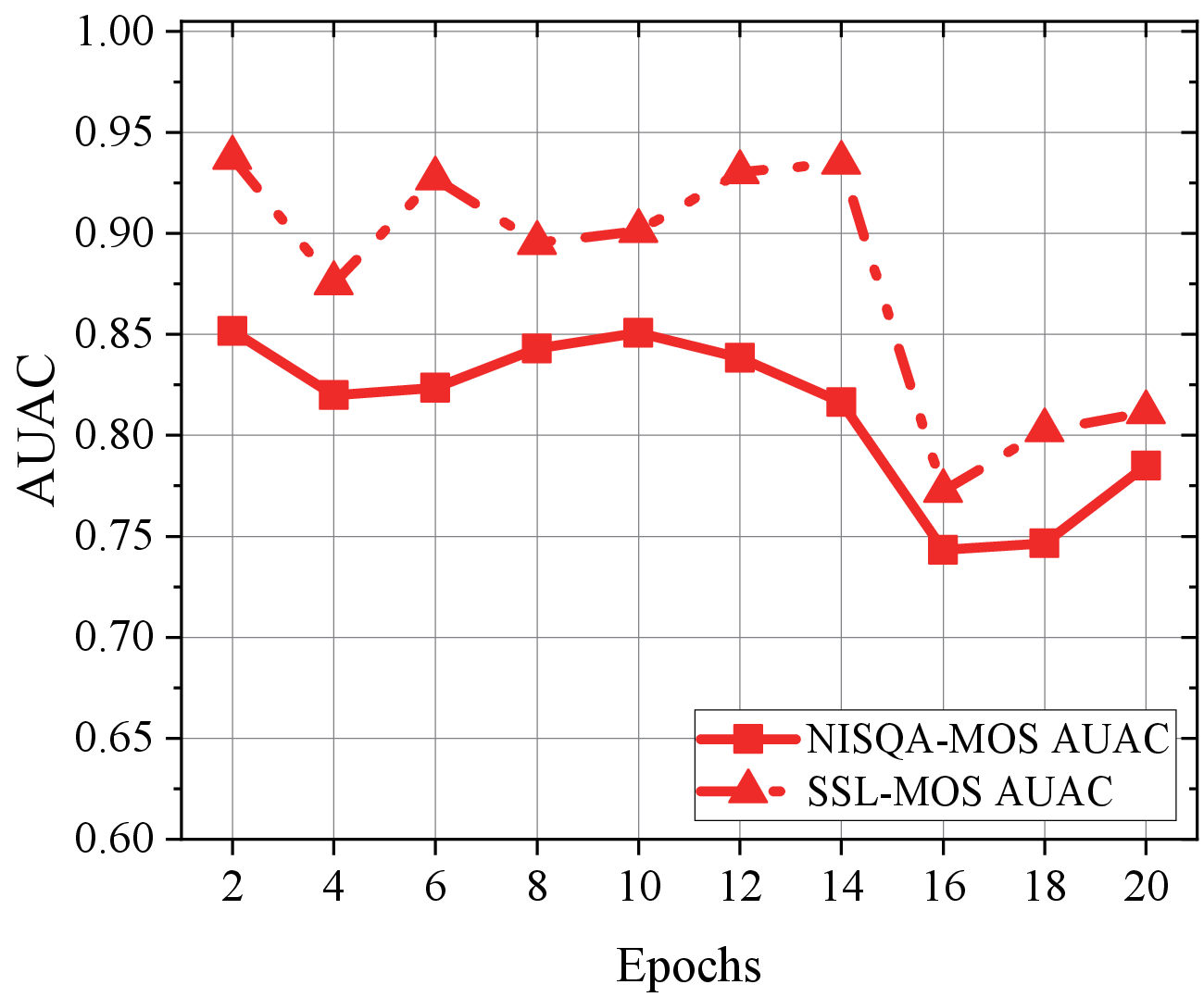}}
  \label{fig:Defense-FineTuning_Tencent}
  \subfloat[VCC 2018 Dataset]{
    \includegraphics[width=0.43\linewidth]{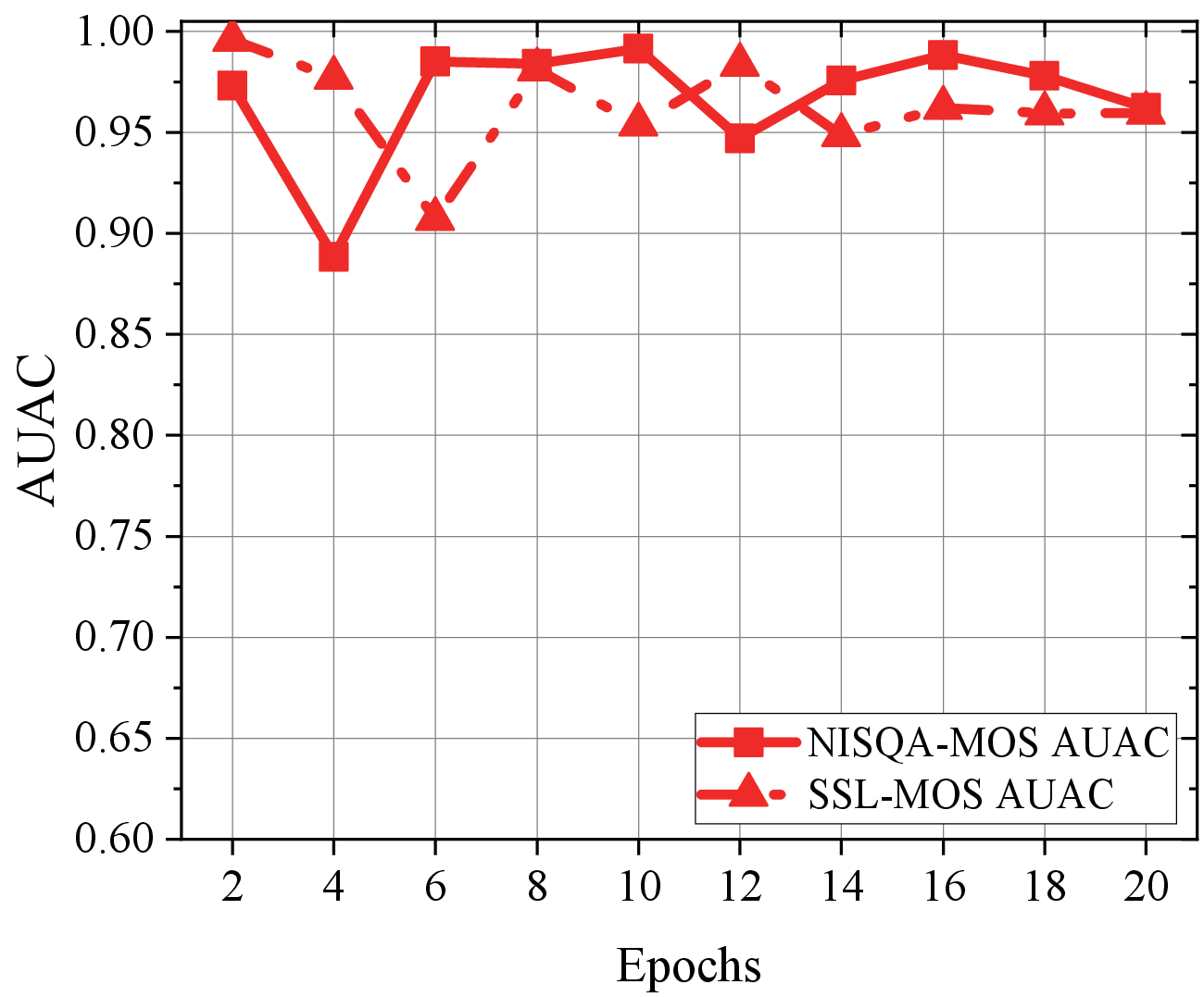}}
  \label{fig:Defense-FineTuning_VCC}
  \subfloat[SOMOS Dataset]{
    \includegraphics[width=0.43\linewidth]{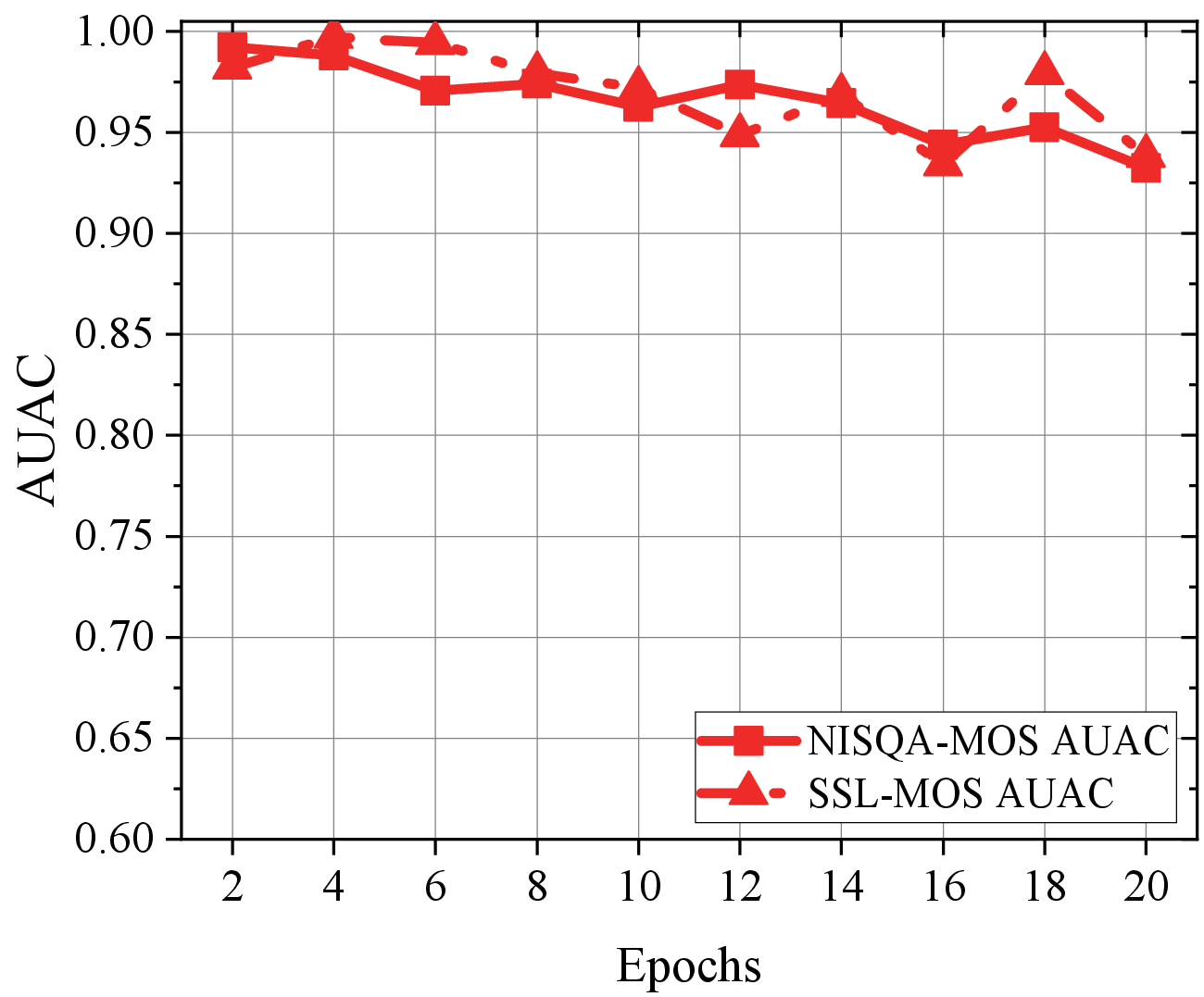}}
  \label{fig:Defense-FineTuning_SOMOS}\hfill
  \caption{The resistance of the model against fine-tuning.}
    \label{fig:Defense-FineTuning}
\end{figure}

\textbf{Resistance to Fine-Tuning.}
We conducted fine-tuning on the model using new data to evaluate EventTrojan's resistance to it. As shown in Fig.~\ref{fig:Defense-FineTuning}, EventTrojan has good resistance to fine-tuning, being affected more only on Tencent Corpus. The AUAC is as low as 0.74 on this dataset, but the RASR at this point is 55.16\%, still an acceptable attack performance.

\section{Conclusion}
\label{sec:conclusion}
This work represents an in-depth exploration of the NISQA model against backdoor attacks. Our EventTrojan uses specific events as triggers to successfully launch backdoor attacks with both a high attack success rate and high stealth. Moreover, the scenarios are from real-life online meeting and voice conversion scenarios, which are highly threatening. Through this approach, EventTrojan emphasizes the urgent necessity for robust backdoor defenses in NISQA for security-critical applications. In addition, we provide a subjective and equitable metric for backdoor attacks on regression tasks to measure attack performance. We hope this paper will draw more researchers' attention to NISQA security issues and promote the establishment of trustworthy NISQA systems.

\section{acknowledgement}
\label{acknowledgement}
This work was supported by the National Natural Science Foundation of China (Grant No. 62171244, 61901237), Ningbo Science and Technology Innovation Project (Grant No. 2022Z074, 2022Z075, 2023Z067), Industry-University-Research Innovation Fund of Chinese Universities (Grant No. 2022MU063) and Key Laboratory of Computing Power Network and Information Security, Ministry of Education, Qilu University of Technology (Grant No. 2023ZD026).
\bibliographystyle{IEEEbib}
\bibliography{conference_101719}

\end{document}